\documentclass{jps-cp}
\usepackage{graphicx}
\usepackage{wrapfig}
\usepackage{txfonts} % Comment out this line unless the txfonts package is available in your LaTeX system.

\title{Slave-Rotor and Auxiliary Chain Analysis of the Mott Transition in a Modified Periodic Anderson Model}

\author{Ankur \textsc{Majumder}$^{1}$ and Sudeshna \textsc{Sen}$^{1}$}

\inst{$^{1}$Department of Physics, Indian Institute of Technology (Indian School of Mines), Dhanbad, 826004, India}

\email{22dr0052@iitism.ac.in, sudeshna@iitism.ac.in}

\recdate{\today}
\abst{
The Mott-Hubbard transition is a paradigmatic phenomenon driven by strong electronic correlations. We investigate this transition in a three-orbital modified periodic Anderson model using slave-rotor mean-field theory. A minimal two-site cluster extension is introduced to understand the minimal effects of short-range spatial correlations. In comparison to the single-site solution, short-range spatial correlations reveal a shift of the transition point. Going beyond the slave-rotor analysis, we also evaluate the dynamical self-energy and map it onto the boundary Green's function of a non-interacting semi-infinite auxiliary chain. For the metallic phase, the resulting hopping pattern corresponds to a topologically trivial generalized Su--Schrieffer--Heeger chain. This construction provides a framework for exploring the possible topological character of the Mott transition in this model.}

\kword{Mott metal-insulator transition, slave-rotor formalism, cluster mean-field theory, periodic Anderson model, dynamical mean-field theory}

\begin{document}
\raggedbottom
\maketitle

% \section{Introduction}

% You can use this file as a template to prepare your manuscript for JPS Conference Proceedings\cite{cp,jpsj,ptep,instructions,format}.

\section{\label{sec:level1}Introduction}

The Mott-Hubbard metal-insulator transition (MIT) is a central phenomenon in strongly correlated electron systems~\cite{Mott_review,mott3,mott1961transition} that occurs when electron-electron repulsion competes with itinerancy, opening a charge gap and invalidating the independent-electron picture. Its non-perturbative character is also central to other correlation phenomena, including quantum criticality and the Kondo effect~\cite{paschen2021quantum,hewson1993kondo}. 
%Unlike a band insulator, whose gap follows from one-electron band structure~\cite{ashcroft2022solid}, a Mott insulator requires a many-body description. 
The single-band Hubbard model is the canonical minimal model~\cite{mott1949basis,gonzalez1995mott}; exact results are available in one dimension and in the infinite-dimensional, or infinite-coordination, limit~\cite{lieb1968absence,vollhardt2022julich}. Dynamical mean-field theory (DMFT) provides a non-perturbative treatment of such systems and becomes exact in the infinite-coordination limit~\cite{dmft_rmp,metzner_vollhardt_infinite_dim}. Applications to Hubbard and periodic Anderson models have clarified the physics of Mott transitions, heavy fermions, and Hund's metals~\cite{Phys_today_DMFT}; combined with density-functional theory, DMFT also provides a predictive framework for correlated materials~\cite{paul2019applications,biermann2006electronic}.

Here we study the modified periodic Anderson model, a minimal three-orbital, equivalently a Kondo-insulating layer, coupled to a metallic layer. Previous DMFT studies found a Mott transition with quantum-critical signatures at zero temperature~\cite{sen2016quantum,sen2023bilayer,sujan2025}. Because single-site DMFT neglects spatial correlations, cluster DMFT is required for non-local effects but becomes computationally expensive as the cluster grows. Following our earlier two-site cluster study~\cite{ankur2025}, we develop a semi-analytic two-site extension of slave-rotor mean-field theory to estimate short-range correlation effects. Going beyond the slave-rotor mean field theory, we analyze the real-frequency DMFT self-energy evaluated using the local moment approach~\cite{LMA_hbar} and numerical renormalisation group as the impurity solvers\cite{bulla2008numerical}.  Motivated by the topological characterization of the Hubbard-model Mott transition~\cite{sen2020Motttopology}, we adopt the auxiliary chain representation to explore the topological interpretation of the phases in this model. %We focus on the Fermi-liquid phase and discuss how the construction may be extended to the transition and insulating regimes. 

\section{\label{sec:model_formalism}Model}

\begin{figure}[htp!]
\centering
\includegraphics[width=0.35\linewidth]{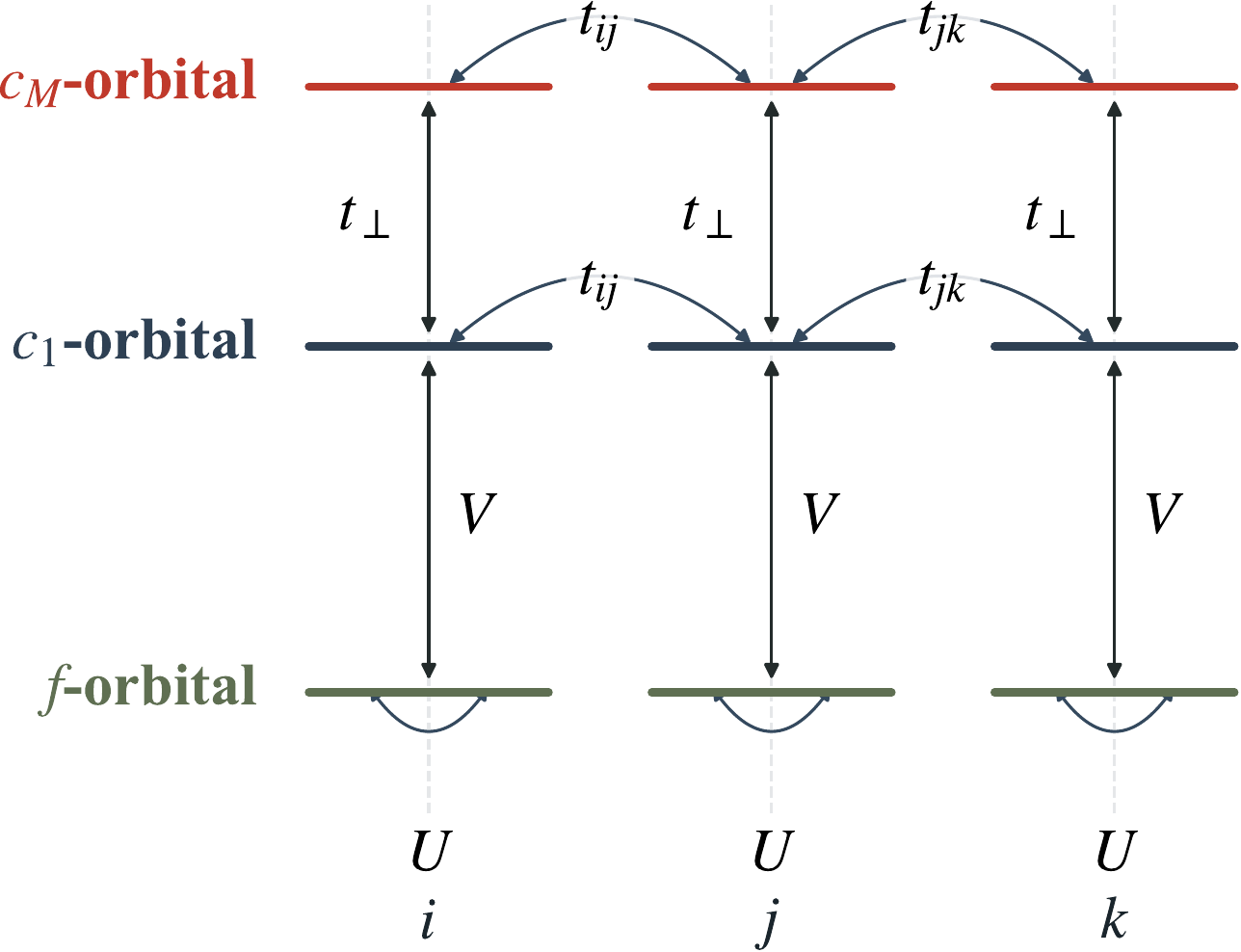}%
%\hfill%
\includegraphics[width=0.325\linewidth]{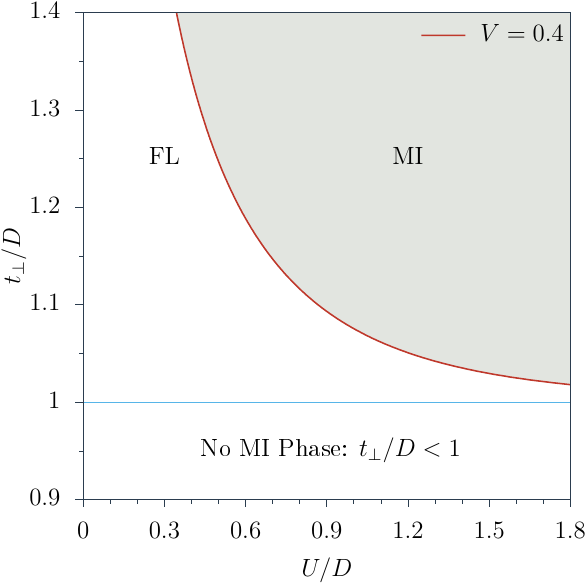}
\caption{\label{fig1} (left)  A schematic of the model Hamiltonian studied in this work. Each site ($i$) on the lattice consists of three orbitals: a highly localized $f$ orbital that hybridizes with a delocalized $c$ orbital via a hybridization energy $V$. The $c$ orbital is further coupled to a second set of delocalized orbitals ($c_M$) via the hopping integral $t_\perp$. The electrons in the delocalized $c$ and $c_M$ orbitals hop around intra-orbitally with hopping energy $t$ between nearest neighbors and are non-interacting. The electrons on the localized $f$ orbitals interact with each other via a repulsive (Hubbard) interaction $U$. (right) Zero-temperature two-site DMFT phase diagram of this model, where the red line separates the Fermi liquid (FL) and Mott (MI) states, and there is no MI phase for $t_\perp/D<1$.}
\end{figure}

% \begin{wrapfigure}[20]{r}[2pt]{0.4\textwidth}
% \vspace{-0.7\baselineskip}
% \centering
% \includegraphics[width=1\linewidth]{fig_1.pdf}\\[-0.2em]
% \includegraphics[width=0.95\linewidth]{fig_2.pdf}
% \caption{\label{fig1} (top)  A schematic of the model Hamiltonian studied in this work. Each site ($i$) on the lattice consists of three orbitals: a highly localised $f$ orbital (depicted as the blue layer) that hybridises with a delocalised $c$ orbital (yellow layer) via a hybridization energy $V$. The $c$ orbital is further coupled to a second set of delocalised orbital ($c_M$, brown layer) via hopping integral $t_\perp$. The electrons in the delocalised $c$ and $c_M$ orbitals hop around intra-orbital with hopping energy $t$ between nearest neighbours and are non-interacting. The electrons on the localised $f$ orbitals interact with each other via a repulsive (Hubbard) interaction $U$. (bottom) Zero-temperature two-site DMFT phase diagram, where the red line separtaes the Fermi liquid (FL) and Mott(MI) states.}
% \vspace{-0.7\baselineskip}
% \end{wrapfigure}
The modified periodic Anderson Hamiltonian contains one interacting, localized $f$ orbital and two non-interacting, itinerant orbitals, $c$ and $c_M$, per lattice site. Electrons hop between nearest-neighbor sites with amplitude $t$; the $f$ and $c$ orbitals hybridize locally with amplitude $V$, while $c$ and $c_M$ are coupled by $t_\perp$. The Hamiltonian is $H=H_f+H_c+H_{cM}+H_{\rm hyb}$, where,

\begin{align}
H_f+H_c+H_{cM}={}&\epsilon_f\sum_{i\sigma}f^\dagger_{i\sigma}f_{i\sigma}
+\frac{U}{2}\sum_i\left(\sum_\sigma n_{fi\sigma}-\frac{1}{2}\right)^2 -t\sum_{\langle ij\rangle\sigma}\left(c^\dagger_{i\sigma}c_{j\sigma}
+c^\dagger_{Mi\sigma}c_{Mj\sigma}+\mathrm{h.c.}\right),\\
H_{\rm hyb}={}&V\sum_{i\sigma}\left(f^\dagger_{i\sigma}c_{i\sigma}+\mathrm{h.c.}\right)
+t_\perp\sum_{i\sigma}\left(c^\dagger_{i\sigma}c_{Mi\sigma}+\mathrm{h.c.}\right).
\end{align}

Here, $\sigma=\uparrow,\downarrow$ labels spin. For $t_\perp=0$, the model reduces to the periodic Anderson model, whose particle-hole-symmetric half-filled state ($n_f+n_c=2$) is a Kondo insulator. The full model may therefore be viewed as a Kondo-insulating layer coupled to a metallic layer. We set $\epsilon_f=\epsilon_c=\epsilon_{cM}=0$ to impose particle-hole symmetry.

At fixed $V$, varying either $U$ or $t_\perp$ drives a Mott-Hubbard MIT~\cite{sen2016quantum,sen2023bilayer,ankur2025}. The phase boundary in Fig.~\ref{fig1} shows that the Mott phase exists only for $t_\perp/D\geq1$, where $D$ is the conduction-electron half-bandwidth~\cite{ankur2025}. We next apply slave-rotor mean-field theory~\cite{Florens_2004,Florens_2002} and construct a two-site cluster extension to estimate non-local effects~\cite{Paramekanti_2007}.

\section{Slave rotor mean-field formalism for modified periodic Anderson Model}

\begin{figure}[!t]
\centering
\includegraphics[width=0.6\linewidth]{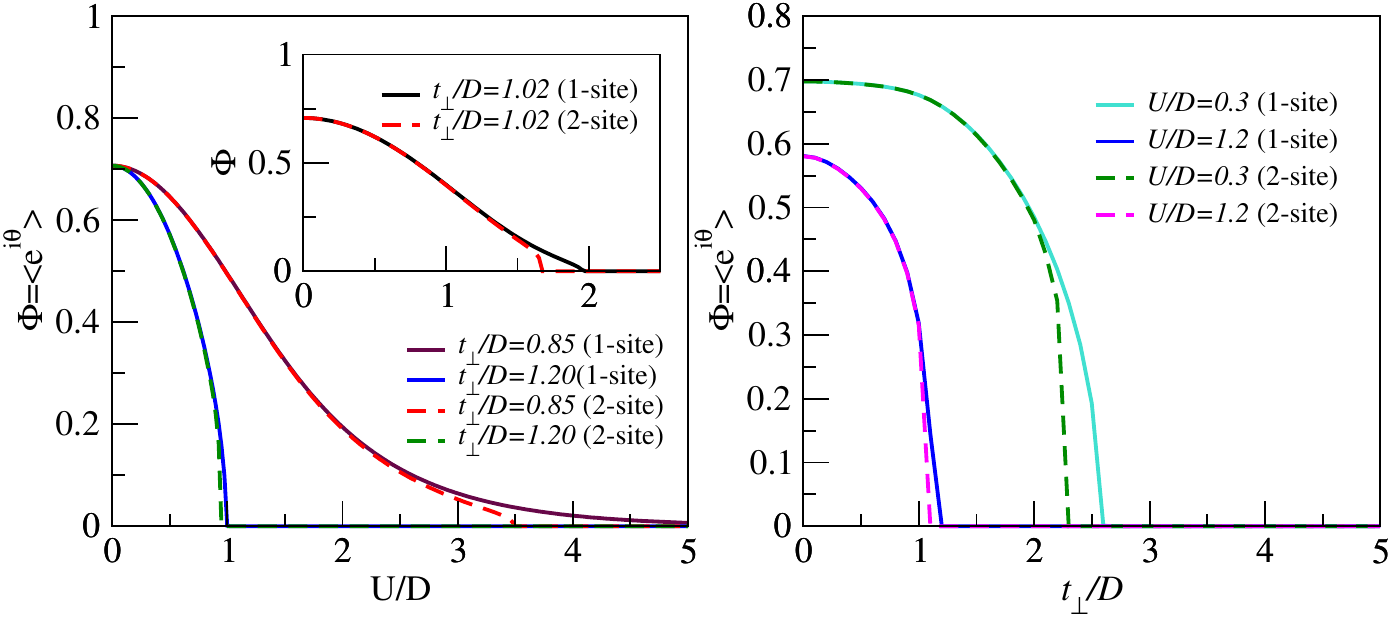}%
\caption{Slave-rotor quantification of quasi-particle weight, $\Phi=\langle e^{-i\theta}\rangle$, comparing the single-site (solid lines) and two-site cluster (dashed lines) mean-field results. Left: $\Phi$ as a function of $U/D$ for $t_\perp/D=0.85$ and $1.20$; the inset shows the comparison near the phase boundary at $t_\perp/D=1.02$. Right: $\Phi$ as a function of $t_\perp/D$ for $U/D=0.3$ and $1.2$. The vanishing of $\Phi$ signals Mott localization, while the shift between the single-site and cluster results gives a qualitative estimate of the effect of short-range spatial correlations on the transition.}
\label{fig:slave_rot}
\end{figure}

In this section, we treat the Hamiltonian above using the standard slave-rotor formalism~\cite{Florens_2004}. In this theory, original \textit{f}- electrons are written as product of pseudo-fermionic \textit{d}-operators (called, s spinon) and slave-rotor \textit{$e^{\pm i\theta}$}-operators (called, chargons), given by the transformation equation: $f_{i\sigma}^\dagger(f_{i\sigma})=d_{i\sigma}^\dagger e^{i\theta_i} (d_{i\sigma}e^{-i\theta_i})$, where the $e^{i\theta}$ ($e^{-i\theta}$) represents the raising (lowering) operator in the angular-momentum basis. These transformation equations are subjected to the constraint $L_i=\sum_\sigma n^f_{i\sigma}-1/2=\sum_\sigma n^d_{i\sigma}-1/2$ to restrict the unphysical states; here, $n^{d(f)}$ represents the number operator corresponding to the $d(f)$ fermions. Thus, the transformed Hamiltonian, constrained via the Lagrange multiplier $\lambda$, is
\WFclear
\begin{align}
	H_{\rm SR} &=\epsilon_f\sum_{j\sigma}n^d_{j\sigma}+\frac{U}{2}\sum_{i}\left(L_{i}\right)^2+\lambda\sum_{i\sigma}\left(L_{i}+\frac{1}{2}-n^d_{i\sigma}\right)-t\sum_{<ij>\sigma}\left( c_{i\sigma}^\dagger c_{{j\sigma}} + h.c\right) -t\sum_{<ij>\sigma}\left( c_{M_{i\sigma}}^\dagger c_{M_{j\sigma}} +h.c\right)\nonumber\\&+t_\perp\sum_{i\sigma}\left(c_{i\sigma}^\dagger c_{M_{i\sigma}} + h.c\right)+V\sum_{i\sigma}\left( c_{i\sigma}^\dagger d_{i\sigma}e^{-i\theta_i} + h.c\right) \nonumber
\end{align}
Now, $H_{\rm SR}$ is decoupled using mean-field theory as $H_{\rm SR}=H_{\theta}+H_d$, where $H_{\theta}$ and $H_d$ are the chargon and spinon Hamiltonians, respectively. This is done by subjecting the only spinon-chargon coupling term present in $H_{\rm SR}$, to mean-field treatment as,$\left<c_{i\sigma}d_{i\sigma}e^{-i\theta}\right>_{MF}=\left<c_{i\sigma}d_{i\sigma}\right>_de^{-i\theta}+\left<e^{-i\theta}\right>_\theta c_{i\sigma}d_{i\sigma}$, where, $\left<...\right>_\theta$ and $\left<...\right>_d$ means expectation with respect to chargon and spinon states respectively. The decoupled Hamiltonian is given as,
\begin{align}
	H_d=&-t\sum_{<ij>\sigma}\left( c_{i\sigma}^\dagger c_{{j\sigma}} + h.c\right) -t\sum_{<ij>\sigma}\left(  c_{M_{i\sigma}}^\dagger c_{M_{j\sigma}}+ h.c\right)+t_\perp\sum_{i\sigma}\left(c_{i\sigma}^\dagger c_{M_{i\sigma}} + h.c\right)\nonumber \\&+\left(\epsilon_o-\lambda\right)\sum_{i\sigma}n^d_{i\sigma}+ V_{\rm eff}\sum_{i\sigma}\left( c_{i\sigma}^\dagger d_{i\sigma} + h.c\right),  \label{eq:pfer}\\
	H_\theta=&\frac{U}{2}\sum_{i}\left(L_{i}\right)^2+\lambda\sum_{i}L_{i}+\sum_{i\sigma} h_i e^{-i\theta_i}+h.c. .\label{eq:rot}
\end{align}
where, $V_{\rm eff}=V\left<e^{-i\theta_i}\right>_\theta$ and $h=V\left<c_{i\sigma}d_{i\sigma}\right>_d$. The pseudo fermionic part, $H_d$, can be expressed in Fourier space as $H_d=\sum_{k\sigma}\bigl[(\epsilon_f-\lambda)d^\dagger_{k\sigma}d_{k\sigma}+\allowbreak V_{\rm eff}(c_{k\sigma}^\dagger d_{k\sigma}+h.c.)+\allowbreak t_\perp(c_{k\sigma}^\dagger c_{M,k\sigma}+h.c.)+\allowbreak \epsilon_k c_{k\sigma}^\dagger c_{k\sigma}+\allowbreak \epsilon_k c_{M,k\sigma}^\dagger c_{M,k\sigma}\bigr]$, where $\epsilon_k=-\sum_jt_{ij}e^{(\vec{r}_i-\vec{r}_j)\cdot\vec{k}}$ is the energy dispersion of the $c$ and $c_M$ orbitals. For, single-site approximation, only one site is taken, in the presence of the rest of the site as mean-field bath, transforming Equation~\ref{eq:rot} to, $H_{\theta,\textrm{1-site}}=\frac{U}{2}\left(L\right)^2+\lambda L+ 2 (h e^{-i\theta}+h.c.)$. where, 2 factor comes because of the spin-degeneracy and $h$-parameter as calculated from $H_d$ is given by,
\begin{equation}
	h=\frac{1}{\pi\left<e^{-i\theta}\right>}\int_{\omega}\Im\left[\int_{\epsilon}\frac{F_{V\tau}(\omega,\epsilon)D(\epsilon)}{\omega^+-\left(\epsilon_f-\lambda\right)-F_{V\tau}(\omega^+,\epsilon)}d\epsilon\right]d\omega\label{eq:h_fac}
\end{equation}
where $F_{V\tau}(\omega^+,{\epsilon})=V_{\rm eff}^2/\left(\omega^+-\epsilon-\frac{t_\perp^2}{\omega^+-{\epsilon}}\right)$ and $D(\epsilon)=\sum_k\delta(\epsilon-\epsilon_k)$ is the density of state for delocalized orbitals. In the next subsection, we discuss the changes needed to incorporate non-local effects using cluster mean-field theory.

% \subsection{Extension to cluster mean-field theory}

The cluster mean-field theory is an extension of the single-site mean field theory, where we take on a finite number (instead of a single site) of sites (in our case, we are using two sites) in the presence of the other sites via a mean field bath.
Therefore, the 2-site slave-rotor cluster version of the rotor-Hamiltonian, Eq.~\ref{eq:rot} is given by, $H_\theta=\frac{U}{2}\left(L_{1}\right)^2+\frac{U}{2}\left(L_{2}\right)^2+\lambda (L_{1}+ L_{2}) + 2  \left(h_1 e^{-i\theta_1}+ h_2 e^{-i\theta_2}+ h.c. \right)=\sum_{a=1,2} \frac{U}{2}L_{a}^2+\lambda L_{i} + 2  \left(h_i e^{-i\theta_i}+ h.c. \right) $. In Figure~\ref{fig:slave_rot}, we compare the single-site and two-site cluster results. In the left panel, $\Phi$ (related to quasi-particle weight $Z=|\Phi|^2$) decreases with increasing $U$ and vanishes at a finite interaction for $t_\perp/D>1$, marking the Mott transition; the inset highlights the cluster-induced shift near $t_\perp/D=1.02$. The right panel shows the analogous loss of rotor coherence as $t_\perp$ increases, with $U$ held fixed. The cluster correction is largest at smaller $t_\perp$ for the $U$-driven transition and at smaller $U$ for the $t_\perp$-driven transition.

\section{\label{sec:aux_chain}Mapping onto a non-interacting auxiliary chain}

\begin{figure}[t!]
\centering
\includegraphics[width=0.9\linewidth]{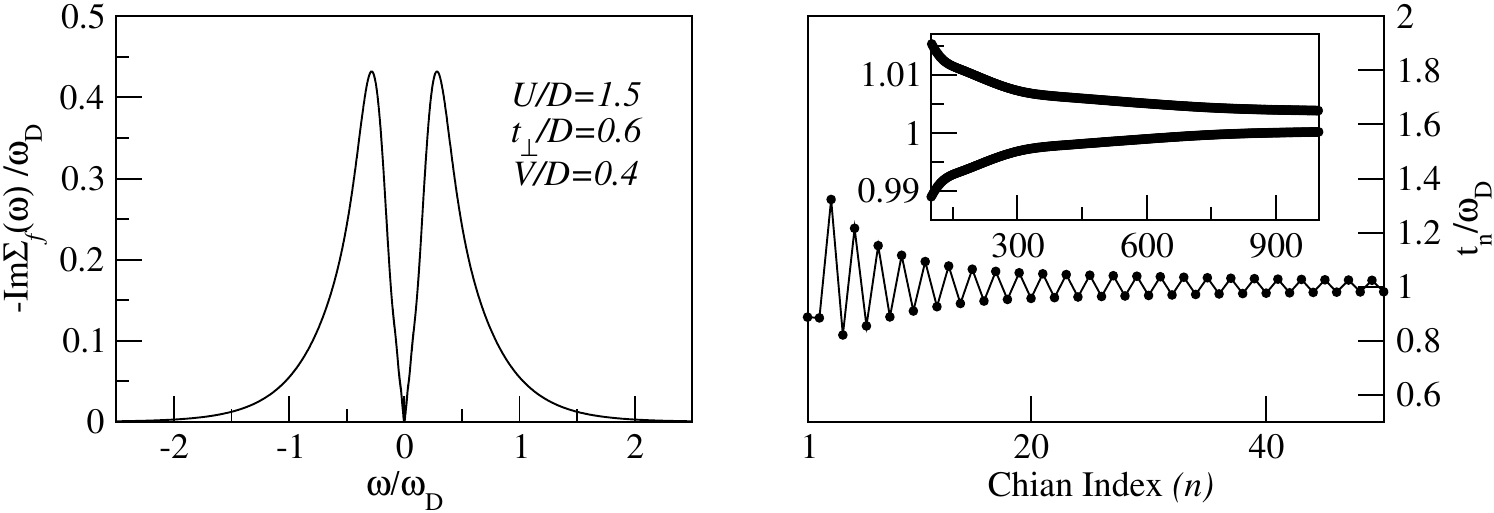}%
\caption{Left: DMFT $f$-electron self-energy for $t_\perp/D=0.6$, $U/D=1.5$, and $V/D=0.4$, 
showing $-\mathrm{Im}\,\Sigma_f(\omega\to 0)\sim\omega^2$. Right: corresponding auxiliary-chain hoppings. The chain begins with a strong bond, $t_{2n-1}/t_{2n}>1$, while the even and odd branches approach one another with a decaying envelope.}
\label{fig:FL}
\end{figure}

\begin{figure}[hbtp!]
\centering
\includegraphics[width=0.9\linewidth]{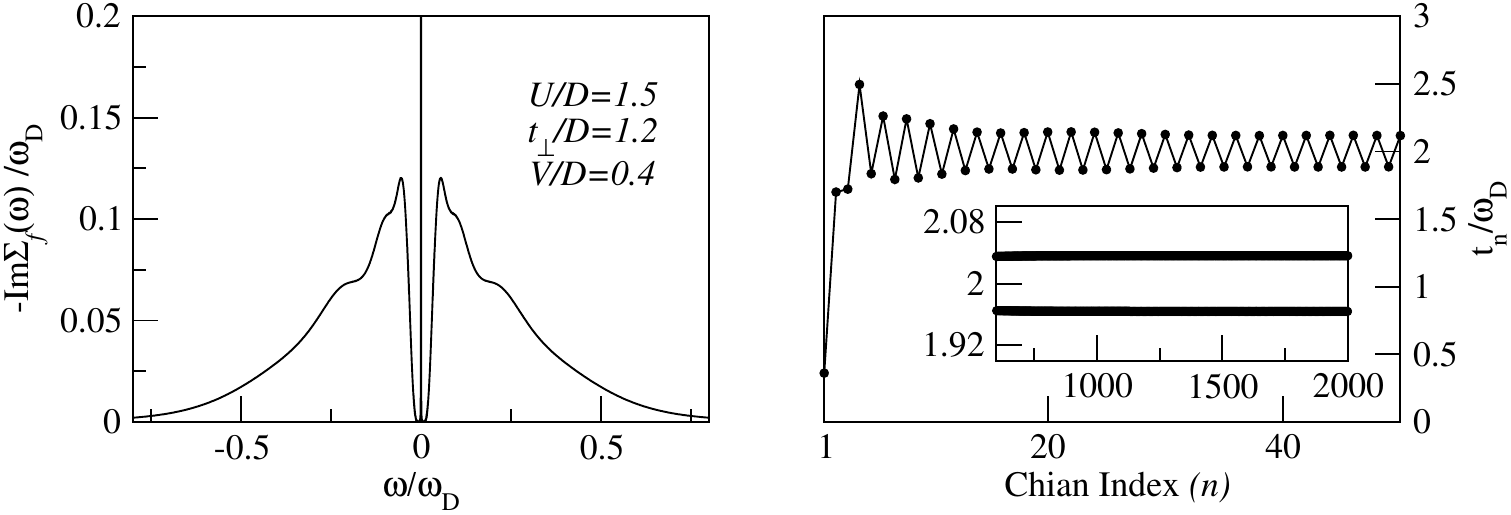}%
\caption{Left: DMFT $f$-electron self-energy for $t_\perp/D=1.2$, $U/D=1.5$, and $V/D=0.4$, 
showing a Mott insulating state. Right: corresponding auxiliary-chain hoppings. The chain begins with a weak bond, resembling a staggered SSH like hopping in the bulk.}
\label{fig:MI}
\end{figure}

The slave-rotor analysis characterises the Mott transition via a vanishing rotor expectation value, but it does not provide us with a dynamical self-energy that embodies the many body dynamics. We evaluate the same using numerical renormalisation group solver within the framework of DMFT, for two representative parameters, namely, $t_\perp/D=0.6$, in the metallic phase and $t_\perp/D=1.2$, in the Mott insulating phase. The interaction strength and the hybridization energy is fixed at $U/D=1.5$, and $V/D=0.4$, respectively. These are plotted in the left panels of Fig.~\ref{fig:FL} (for the metallic phase) and Fig.~\ref{fig:MI} (for the insulating phase). 
%\begin{figure}[!t]
%\centering
%\includegraphics[width=0.43\linewidth]{fig_3_a.eps}%
%\hfill%
%\includegraphics[width=0.43\linewidth]{fig_3_b.eps}
%\caption{Left: DMFT $f$-electron self-energy for $t_\perp/D=0.6$, $U/D=1.75$, and $V/D=0.44$, showing $-\mathrm{Im}\,\Sigma_f(\omega)\sim\omega^2$. Right: corresponding auxiliary-chain hoppings. The chain begins with a strong bond, $t_{2n-1}/t_{2n}>1$, while the even and odd branches approach one another with a decaying envelope.}
%\label{fig2}
%\end{figure}

%Motivated by the recent work on the topological characterization of the Mott transition in the Hubbard model~\cite{sen2020Motttopology}, we analyze the self-energy obtained in this model for a similar characterization. 
%In this work, we demonstrate the study for a single parameter in the Hamiltonian, namely, $t_\perp/D=0.6$, $U/D=1.75$, and $V/D=0.44$. The respective self-energy is plotted in Fig.~\ref{fig2}(left). Note that this is the self-energy obtained by a full DMFT calculation (beyond the linearized DMFT) using a high-precision impurity solver called the Local Moment approach~\cite{sen2016quantum,logan1998local}. 
This self-energy is then mapped onto a non-interacting tight-binding model. The reader is directed to the Reference~\cite{sen2020Motttopology} for more details on this representation. The chain is of a generalized Su-Schrieffer-Heeger (SSH) type chain~\cite{ssh,shortcourse}, where the physical degrees of freedom are coupled to the auxiliary chain via a hopping integral determined by spectral sum rules. It is the boundary site of the auxiliary chain shown in Figure~\ref{fig:FL}(right) that encodes the topological edge state if present. It turns out that the typical $\omega^2$ pseudo-gap kind of behavior of the imaginary part of the Fermi liquid self-energy yields a trivial SSH chain that starts with a strong bond and decays with a $1/n$ envelope ($n$ being the chain site index). The Mott insulating regime, as shown in Fig.~\ref{fig:MI} shows a perfect dimerization in the bulk, resembling the topological phase of the standard SSH model.

While the above behavior of the auxiliary chain is similar to that obtained in the standard Hubbard model, we realized that one needs a really long chain ($n\sim 1000$ sites) to bring out the decaying envelope and accurately represent the physical degrees of freedom in terms of the auxiliary chain. A more detailed analysis of the entire parameter regime, including the Mott insulating phase and the self-energy near the transition, is left for future work.

\section{\label{sec:conclusions}Conclusions}
We studied the Mott transition in a three-orbital lattice model using slave-rotor mean-field theory and a minimal two-site extension. The quasi-particle weight $Z=\Phi^2$ vanishes in the Mott phase, and the shift between the single-site and two-site results provides a qualitative estimate of short-range correlation effects. This shift is most pronounced at smaller $t_\perp$ for the $U$-driven transition and at smaller $U$ for the $t_\perp$-driven transition. Unlike the corresponding Hubbard-model construction, however, the rotor Hamiltonian factorizes as $H_{\theta,\mathrm{2s}}=\sum_{i=1}^{2}H_\theta^{(i)}$, independently of the structure of $H_d$. Fluctuations mediated by the surrounding lattice are therefore not fully retained, and some non-local physics may be missed. The truncated rotor Hilbert space imposes a second limitation. The exact non-interacting limit requires $Z=\Phi^2=1$ at $U=0$, which the slave-rotor theory recovers only with a sufficiently large angular-momentum basis. Here, each rotor is restricted to $L=0,\pm1$, giving a $3\times3$ two-site chargon basis. Slave-spin theories instead use a filling-dependent gauge parameter to enforce the correct $U=0$ limit~\cite{Medici2017}. The present results are therefore qualitative; quantitative non-local corrections and phase boundaries require a full cluster-DMFT calculation.

Finally, we map the physical self-energy obtained within DMFT onto a non-interacting generalized SSH chain. We focus only on the Fermi-liquid self-energy in this work. However, we envisage unconventional features in the chain as it approaches the transition. We believe that the features would differ from those in the conventional case due to the absence of a clear separation of energy scales. Additionally, it would be interesting to see what happens at finite temperatures, where the self-energy has been shown to take a power-law form at the transition point. We leave these analyses for future study.

\section{\label{sec:acknowledgements}Acknowledgements}
A.M. acknowledges financial support from the Prime Minister's Research Fellowship (PMRF; ID 1602706) and IIT (ISM) R\&D support associated with scholar ID 22DR0052. S.S. acknowledges financial support from the ANRF research project (SRDP/1293/G).

\bibliographystyle{plain} % JPSJ bibliography style file
\bibliography{ref}

% \begin{thebibliography}{9}
% \bibitem{cp} The abbreviation for JPS Conference Proceedings should be ``JPS Conf. Proc." in the reference list.
% \bibitem{jpsj} The abbreviation for the Journal of the Physical Society of Japan should be ``J. Phys. Soc. Jpn." in the reference list.
% \bibitem{ptep} The abbreviation for the Progress of Theoretical and Experimental Physics should be ``Prog. Theor. Exp. Phys." in the reference list.
% \bibitem{instructions} More abbreviations of journal titles are listed in ``Instructions for Preparation of Manuscript", which is available at our Web site (http://jpsj.jps.or.jp).
% \bibitem{format} F. Author, S. Author, and T. Author, Abbreviated journal title \textbf{volume in bold face}, initial page or article number (year of publication).
% \end{thebibliography}

\end{document}